\documentclass[%
 reprint,
aip,
pop,
]{revtex4}

\usepackage{amsmath}
\usepackage{graphicx}% Include figure files
\usepackage{dcolumn}% Align table columns on decimal point
\usepackage{bm}% bold math
\usepackage{overpic}
\usepackage{color}
\usepackage{hyperref}% add hypertext capabilities
\begin{document}

\title{Ponderomotive Injection in Plasma Wakefield Accelerators}% Force line breaks with \\

\author{Ming Zeng}
\email{ming.zeng@desy.de}
\author{Alberto Martinez de la Ossa}
\author{Jens Osterhoff}
\affiliation{
 Deutsches Elektronen-Synchrotron DESY, 22607 Hamburg, Germany
}
\date{\today}% It is always \today, today,

\begin{abstract}
A new electron injection scheme is proposed in sub-relativistic plasma wakefield accelerators. A transverse laser ionizes a dopant gas and ponderomotively accelerates the released electrons in the direction of wake propagation. This process enables electron trapping in the wakefield even for a wakefield potential below the trapping threshold. We study the scheme theoretically and by means of particle-in-cell simulations to demonstrate high-quality beam formation and acceleration with sub-micrometer normalized emittances and sub-percent uncorrelated energy spreads.
\end{abstract}

\maketitle

%\tableofcontents
%\section{\label{sec:intro}Introduction}
Since their invention almost four decades ago, beam-driven plasma-wakefield accelerators\cite{PChenPRL1985} (PWFAs) have undergone a tremendous development and, today, are regarded as an enticing technology for the next generation of compact beam sources for photon science applications and high-energy physics. Strong electron-density wakes in plasma support electric fields on the order of $E_0 = k_p m_e c^2/e \approx 9.6\times \sqrt{n_p[10^{16}\ \rm cm^{-3}]}\ {\rm GV/m}$, where $k_p=\sqrt{4\pi r_e n_p}$ is the plasma wavenumber, $r_e$ is the classical electron radius, $n_p$ is the plasma density, $c$ is the speed of light in vacuum, $m_e$ is the electron mass, and $e$ is the elementary charge. For typical plasma densities ($10^{16} - 10^{18}\ \rm cm^{-3}$), plasma wakefields can thus outperform state-of-the-art radio-frequency (RF) accelerators by 3 to 4 orders of magnitude in acceleration gradient.

In a PWFA, a high current-density drive beam excites density oscillations, the wakefield, in the plasma electron background. A second, co-propagating particle beam, the witness, can extract energy stored in these waves and be accelerated into the direction of the driver. When the particle density of the drive beam exceeds that of the background plasma, wakefields are generated in the non-linear regime, characterized by a cavitation of the electron plasma density referred to as blowout. For sufficiently narrow drivers, the strength of the plasma blowout is determined by the dimensionless parameter $\Lambda = 2 I_{\rm peak} / I_{A}$, where $I_{\rm peak}$ is the driver peak current and $I_A = 17~{\rm kA}$ is the Alfv\'en current. For $\Lambda \gtrsim 1$, the expelled plasma electrons acquire relativistic velocities, and thus this regime can be referred to as the relativistic regime. 
Several schemes have been proposed for the formation of witness electron bunches to be injected into the plasma wake and into an accelerating phase such that they can acquire high energies.   
These injection schemes are based on plasma density transitions~\cite{HSukPRL2001, CGRGeddesPRL2008, AMartinezPRAB2017}, selective ionization of dopant gas species~\cite{MChenJAP2006, EOzPRL2007, APakPRL2010, CMcGuffeyPRL2010, BHiddingPRL2012, AMOssaPRL2013, ADengNatPhys2019} or external magnetic fields~\cite{JVieiraPRL2011}, and often rely on a (near-)relativistic regime to facilitate the trapping of electrons inside the wakefield. If the driver current is well below the relativistic threshold, the plasma wakefield does not have the required strength to trap electrons from rest, thus impeding the application of most injection methods.

\begin{figure}[ht]
	\begin{overpic}[width=0.48\textwidth,trim={15 12 22 28},clip]{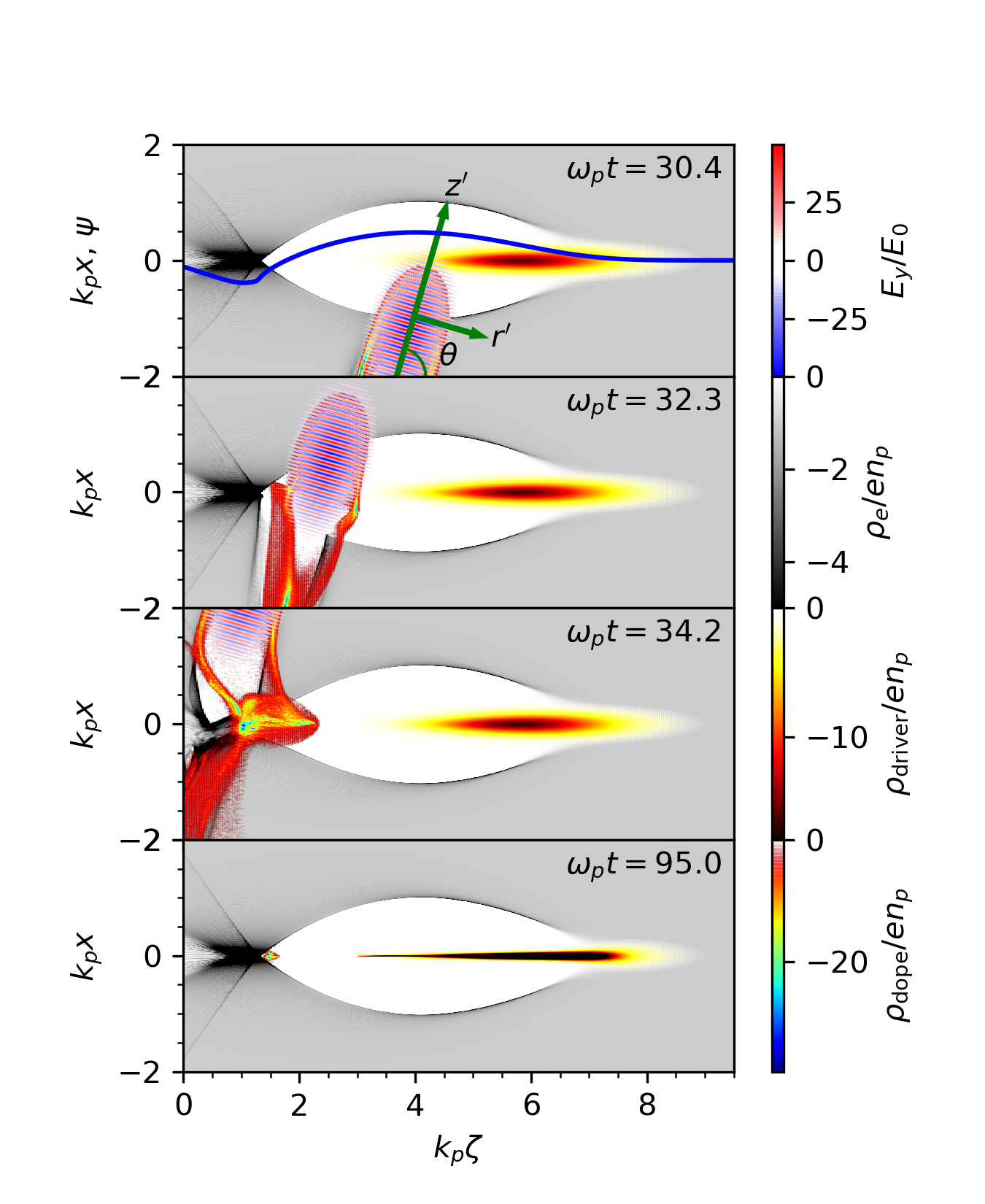}
        	\put(59,75){(a)}
        	\put(59,54){(b)}
        	\put(59,32){(c)}
        	\put(59,11){(d)}
	\end{overpic}
	\caption{Snapshots of a simulation demonstrating ponderomotive injection. The longitudinal axis is translated to a co-moving frame with $\zeta=z-ct$. (a) The injection laser enters from the lower boundary. The blue solid line indicates the axial pseudo-potential $\psi$ of the wake which drops by 0.6 from the peak to the rear of the blow-out, preventing the injection of any electrons from rest. (b) During and (c) after the ponderomotive scattering of electrons ionized from the dopant gas. (d) Electrons gain sufficient momentum in forward direction to get trapped in the wake and form the witness beam.}
	%where electrons from the ionized dopant species are appearing. (d) Further after the injection, when some of the electrons are lost and the core of the electron bunch could remain.}
   \label{fig:illu}
\end{figure}
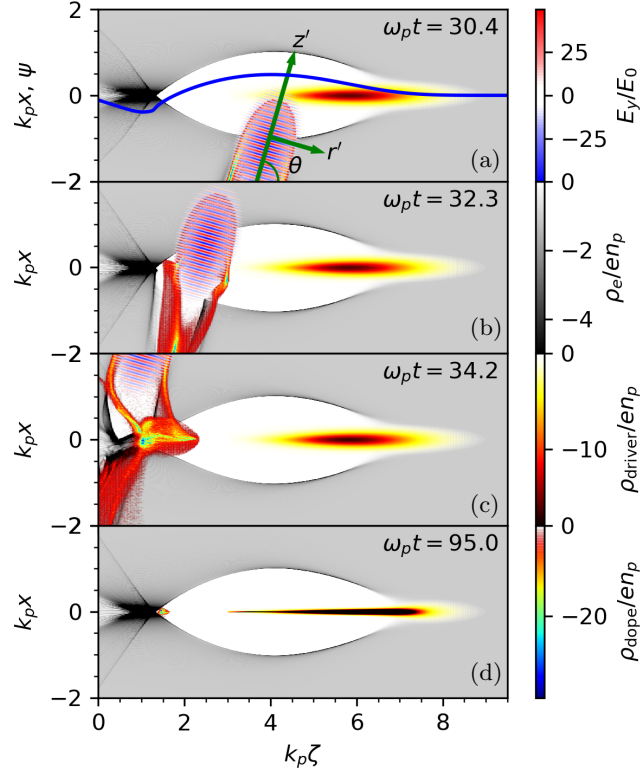

In this Letter, we propose a novel injection scheme for PWFAs operating in the sub-relativistic regime, which utilizes the ponderomotive force of a laser pulse to ease injection of electrons into the plasma wake and form a high-quality witness beam. 
An illustrative example of a full 3 dimensional particle-in-cell (PIC) simulation with the code OSIRIS~\cite{OSIRIS} is shown in Fig.~\ref{fig:illu}. 
A laser with oblique incidence with respect to the direction of propagation of the drive beam ionizes an initially neutral dopant gas (e.g.\ He) co-existing with the background plasma as shown in Fig.~\ref{fig:illu} (a). 
Unlike required for other ionization-based injection mechanisms~\cite{MChenJAP2006, EOzPRL2007, APakPRL2010, CMcGuffeyPRL2010, AMOssaPRL2013}, the wakefield in this scheme is not able to trap any electrons originating at rest from ionization.
Here, the ponderomotive force of the injection laser pre-accelerates a certain fraction of the released electrons and they acquire sufficient momentum in the direction of the drive beam to get trapped in the plasma wake when the laser passes through the blowout cavity, as shown in Fig.~\ref{fig:illu} (b) and (c).
This selective trapping procedure strongly constrains the transverse phase-space volume of trapped electrons in the sub-relativistic wakefield, resulting in electron bunches with potentially low emittance, as shown in Fig.~\ref{fig:illu} (d). The duration of the injection event is constrained by the temporally localized overlap of the laser with the plasma wake, resulting in the generation of witness bunches with small uncorrelated energy spread.

%\section{\label{sec:trap_theory}Trapping theory of pre-accelerated electrons}
In the following, we present a model to describe the dynamics of the electrons from ionization subject to the injection laser and the wakefield and establish a necessary condition for the trapping of a witness bunch. Under the quasi-static approximation, there is a constant of motion for the electrons in a PWFA~\cite{PMoraPOP1997}
\begin{eqnarray}
  \gamma - v_{\phi}p_z - \psi = \text{const.}
  \label{eq:QSAconst},
\end{eqnarray}
where $\gamma = \sqrt{1 + \left|\mathbf{p}_{\perp}\right|^2 + p_z^2}$ is the relativistic factor, $\mathbf{p}=(\mathbf{p}_\perp,\ p_z)$ is the momentum of the electron normalized to $m_ec$, $v_{\phi} \lesssim 1$ is the phase velocity of the wakefield normalized to $c$, $\psi = \varphi - v_{\phi} A_z$ is the normalized pseudo-potential of the wakefield, with $\varphi$ and $\mathbf{A}$ the electric and magnetic potentials normalized to $m_ec^2/e$, respectively.

The trapping condition for an initially stationary electron has been studied well~\cite{WLuPOP2006}. For a non-stationary (pre-accelerated) electron, some differences are expected. Let subscript 0 refer to the initial status of the electron and subscript 1 to the status at the trapping instant. An electron is considered trapped at the instant when it propagates at the same velocity as the wake, i.e.\ $v_{z1} = v_{\phi}$. Thus $\gamma_1 - v_{\phi}p_{z1} = \gamma_{\phi}^{-1}\sqrt{1+\left|\mathbf{p}_{\perp 1}\right|^2}$, where $\gamma_{\phi}\gg 1$ is the relativistic factor of the wake, and usually $\left|\mathbf{p}_{\perp 1}\right| \lesssim 1$. Together with Eq.~(\ref{eq:QSAconst}) this yields the trapping condition
\begin{eqnarray}
  \gamma_0 - p_{z0} = - \Delta \psi + \mathcal{O}\left(\gamma_{\phi}^{-1}\right)
  \label{eq:trap_cond},
\end{eqnarray}
where $-\Delta \psi \equiv \psi_0 - \psi_1 \leq \psi_M$, with $\psi_M$ being the maximum pseudo-potential drop in the wakefield (note $\psi_M > 0$).

In case of a strong driver with $\Lambda \gtrsim 1$ (or $I_{\rm peak} \gtrsim 8.5\ \rm kA$)~\cite{WLuPOP2006}, the generated wakefield, with $\psi_M \gtrsim 1$, allows the trapping of electrons released at rest in positions satisfying  $-\Delta\psi = 1$. This regime is exploited by injection techniques based on the selective ionization of gases, by either the drive beam~\cite{EOzPRL2007}, the wakefields~\cite{AMartinezdelaOssaPOP2015}, or lasers~\cite{BHiddingPRL2012, XLXuPRL2014, FLiPRL2013, MChenPRAB2014}. However, in case of a moderate peak-current driver (such as available e.g.~in the FLASHForward facility at DESY with $I_{\rm peak} \lesssim 3 \rm kA$)~\cite{WAckermannNPho2007, RDArcyRSTA2019}, the maximum achievable $\psi_M<1$~\cite{AMartinezdelaOssaPOP2015}. Thus, electrons can only be trapped, if they acquire a certain initial momentum in longitudinal direction, such that even for a wakefield with $\psi_M < 1$, Eq.~(\ref{eq:trap_cond}) can be fulfilled.

The injection laser is therefore utilized to ionize and provide sufficient forward momentum to the electrons in the right wakefield phase so they satisfy Eq.~(\ref{eq:trap_cond}) and become trapped. To estimate the push of the laser, we assume that the target ionization level of the dopant gas is fully depleted by the very front of the oblique injection laser. According to the ponderomotive model, the effective equation of motion is~\cite{PMoraPOP1997}
\begin{eqnarray}
  \frac{d\mathbf{p}}{dt} = \mathbf{F_{\rm bg}} + \mathbf{F}_{\rm pd}
  \label{eq:e_motion}.
\end{eqnarray}
The first term $\mathbf{F_{\rm bg}}$ is the electromagnetic (EM) force due to the background plasma. The second term
\begin{eqnarray}
  \mathbf{F}_{\rm pd} = -\frac{1}{4\gamma}\mathbf{\nabla} a^2
  \label{eq:pd_force},
\end{eqnarray}
is the ponderomotive force for a linear polarized laser, where $a$ is the normalized profile of the laser vector potential, and
\begin{eqnarray}
  \gamma = \sqrt{1+\left|\mathbf{p}\right|^2+a^2/2}
  \label{eq:gamma_pond},
\end{eqnarray}
is the averaged relativistic factor of the electron.

In order to provide a direct estimate of the maximum momentum achievable during the passage of the laser, we assume that the background EM force is negligible compared to the ponderomotive force. 
By further assuming cylindrical symmetry for the laser profile with respect to its propagation axis, the equation of motion becomes
\begin{eqnarray}
  \frac{dp_{r'}}{dt} = -\frac{1}{4\gamma}\frac{\partial a^2}{\partial r'}
  \label{eq:scatter_motion_r}, \\
  \frac{dp_{z'}}{dt} = -\frac{1}{4\gamma}\frac{\partial a^2}{\partial z'}
  \label{eq:scatter_motion_z},
\end{eqnarray}
where the prime superscripts indicate the laser coordinates. We also assume a simple Gaussian laser profile (close to the focal waist)
\begin{eqnarray}
  a = a_0 \exp\left(-\frac{r'^2}{w^2} -\frac{\zeta'^2}{\tau^2}\right)
  \label{eq:a_profile},
\end{eqnarray}
where $\zeta'=z'-ct'$ is the laser co-moving coordinate. To further simplify, one may re-normalize all the time and length related quantities to $w$ in Eqs.~(\ref{eq:scatter_motion_r})-(\ref{eq:a_profile}), and observe that, for a fixed $a_0$, the final momentum of the electron vs. its initial position $r'_0/w$ does not depend on the absolute value of $w$ or $\tau$, but only on their ratio $\tau/w$.

To obtain the momentum gain after the ponderomotive scattering, we integrate Eqs.~(\ref{eq:scatter_motion_r}) and (\ref{eq:scatter_motion_z}) numerically using the Runge-Kutta 4\textsuperscript{th} order method with the initial conditions of $p_{z'}=p_{r'}=0$, $z'=3\tau$, and varying the initial value of $r'$ in a range (0, $3w$] to model the scattering of the electrons. The integration continues until $p_{r'}$ and $p_{z'}$ do not change anymore, with the final momentum after scattering denoted by $p_{r'\rm sc}$ and $p_{z'\rm sc}$. Finally, we perform axis rotation to obtain the momentum gain in the direction of propagation of the driver
\begin{eqnarray}
  &p_{z0} = p_{z'\rm sc} \cos \theta + p_{r'\rm sc} \sin \theta \cos \varphi \label{eq:pz0_rotation},\\
  &\left|\mathbf{p}_{\perp 0}\right|^2 = \left|\mathbf{p}_{\rm sc}\right|^2 - p_{z0}^2 \label{eq:p_perp_sc},
\end{eqnarray}
where ($\theta$, $\varphi$) are the polar and azimuthal angles of the laser axis with respect to the main wake axis. Eqs.~(\ref{eq:pz0_rotation}) and (\ref{eq:p_perp_sc}) together with Eq.~(\ref{eq:trap_cond}) provide an estimate for the trapping threshold $\psi_{M\rm th} \equiv \gamma_0  - p_{z0}$, which is used to test particle trapping for a certain $\psi_M$.

\begin{figure}
	\begin{overpic}[width=0.48\textwidth]{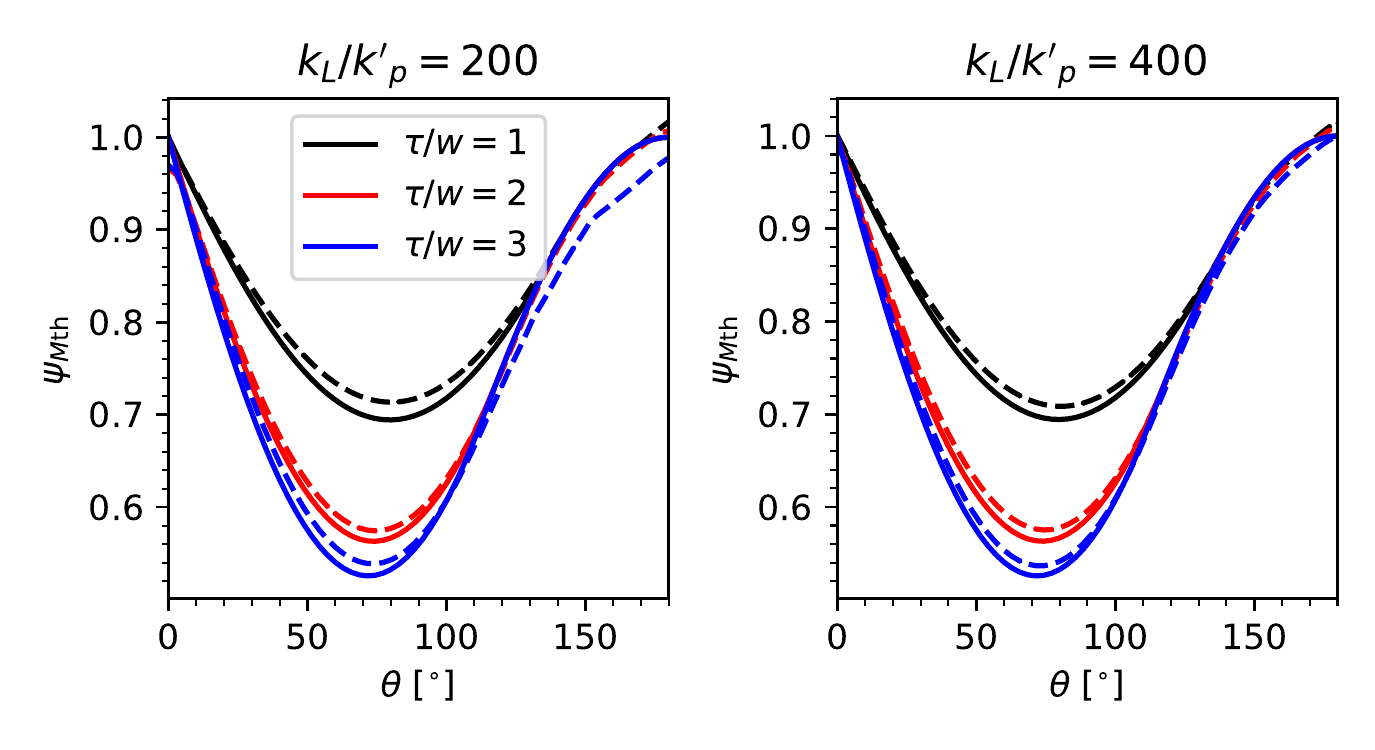}
        	\put(41,14){(a)}
        	\put(90,14){(b)}
	\end{overpic}
	\caption{Verification of the ponderomotive scattering model by PIC simulations. The threshold of $\psi_M$ for trapping to occur ($\psi_{M\rm th}$) vs.\ $\theta$ (the angel between $z$ and $z'$) is plotted for the model (solid curves) and for the PIC simulations (dashed curves). The black, red and blue colors indicate different pulse durations. Two relative frequencies $k_L/k_p' = 200$ (a) and 400 (b) are considered.}
   \label{fig:pds_verify}
\end{figure}

To verify this ponderomotive scattering model, we have performed a series of PIC simulations with negligible wakefield forces compared to the ponderomotive force, i.e.\ when $a/(k_p' w)^2 \gtrsim 1$~\cite{MZengPOP2020}, with $k_p'$ the plasma wavenumber associated with the density $n_{\rm dope}$ of the ionized dopant gas.
We set $k_p' w = 0.1$ and focus the laser at the center of a cubic plasma volume with the size of $\left(0.8/k_p'\right)^3$, such that the length of the interaction region is shorter than the Rayleigh length to prevent a significant change of laser amplitude. We fixed $a_0=1$, varied $\tau/w$ from 1 to 3, and plotted $\psi_{M\rm th}$ vs.\ $\theta$ in Fig.~\ref{fig:pds_verify}, with the solid curves displaying the results from the model and the dashed curves from the PIC simulations, showing good agreement. The scattering is independent of the laser frequency $\omega_L = k_L c$ (where $k_L$ is the laser wave number) for fixed $a_0$ as shown by comparing Fig.~\ref{fig:pds_verify} (a) and (b).

\begin{figure}
	\begin{overpic}[width=0.48\textwidth]{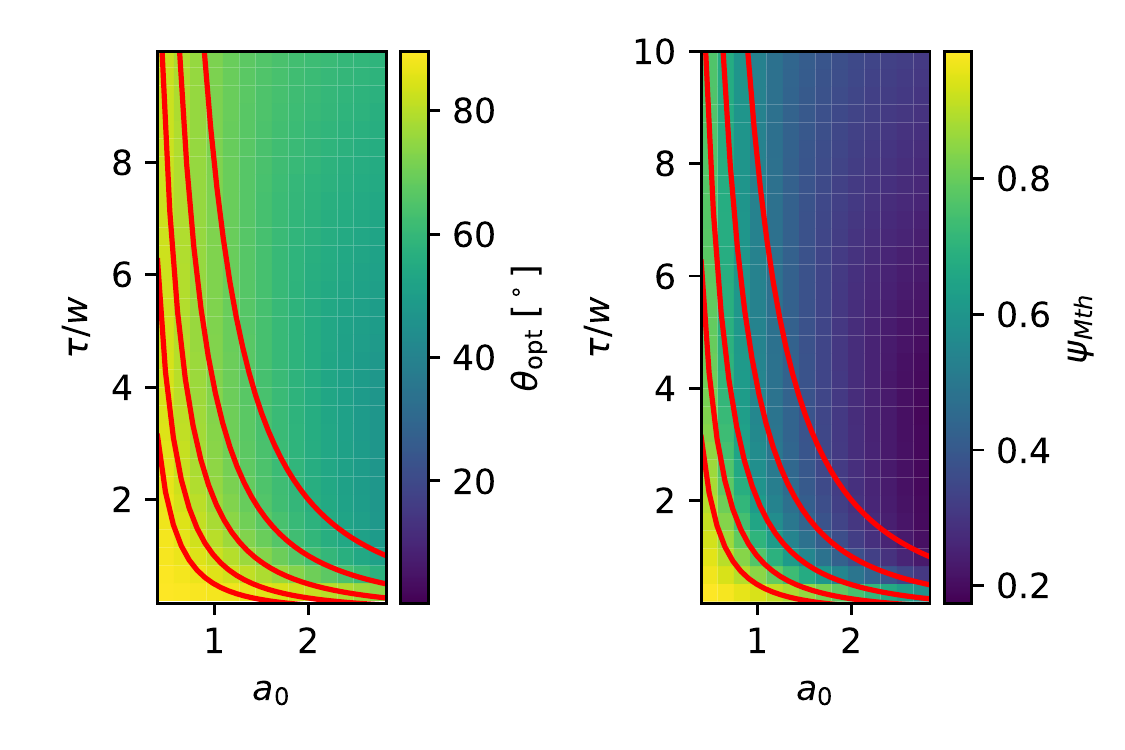}
        	\put(28,57){(a)}
        	\put(76,57){\textcolor{white}{(b)}}
	\end{overpic}
	\begin{overpic}[width=0.48\textwidth]{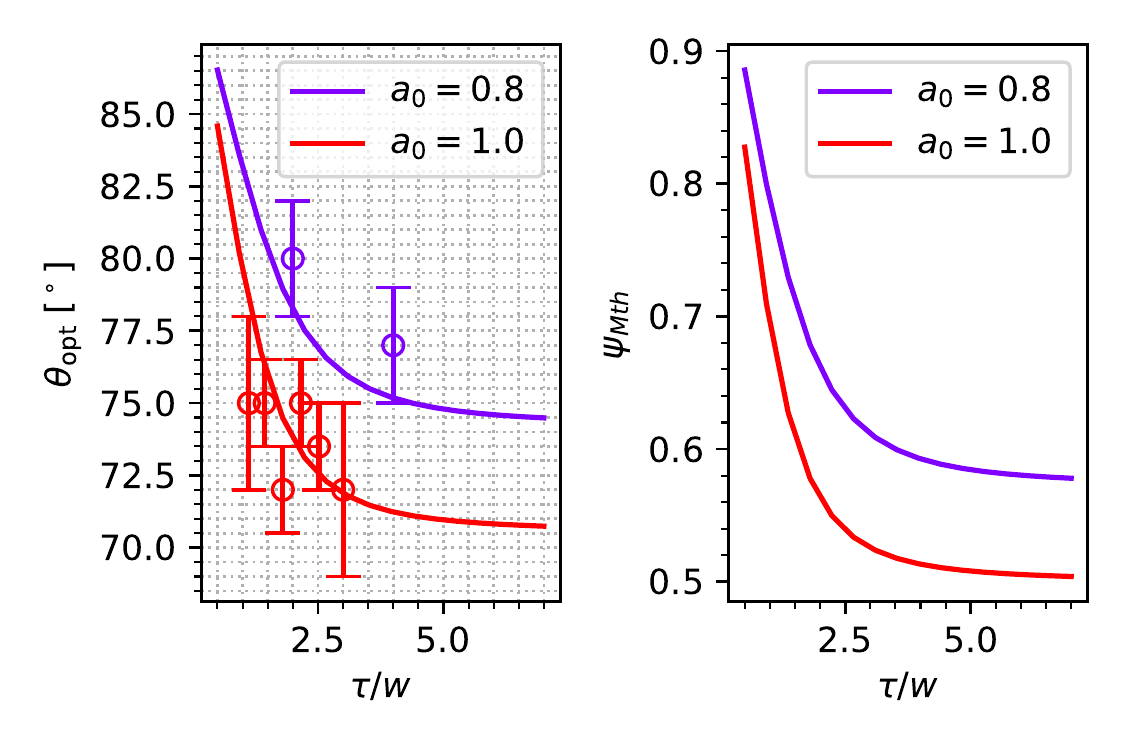}
        	\put(19,15){(c)}
        	\put(66,15){(d)}
	\end{overpic}
	\caption{Scanning of $a_0$ and $\tau/w$ for (a) the optimal angle $\theta$ of the injection laser and (b) the threshold of $\psi_M$ for trapping to occur with optimal $\theta$. The red lines are the contours with the value of $a_0^2 \tau/w$ equaling (left to right) 0.5, 1, 2, 4 and 8. Two values of $a_0 = 0.8$ and 1.0 are shown as lines in (c) and (d). Some PIC simulation scans (with the same driver beam and plasma parameters as in Fig.~\ref{fig:illu}) for the optimal $\theta$ are also shown as circle markers with bars in (c).}
   \label{fig:a0_tau_scan}
\end{figure}

%\section{\label{sec:opt_param}Optimal Scattering Parameters}
Using this ponderomotive scattering model, we scanned $a_0$ and $\tau/w$ to find the optimal angle $\theta_{\rm opt}$ which minimizes $\psi_{M\rm th}$. Fig.~\ref{fig:a0_tau_scan} (a) and (b) show $\theta_{\rm opt}$ and $\psi_{M\rm th}$ as pseudocolors and the contours with equal values of $a_0^2 \tau/w$ as solid lines. Since $w$ is fixed, the contours of $a_0^2 \tau/w$ have fixed laser beam energy. From Fig.~\ref{fig:a0_tau_scan} (b) one can conclude that a direct way to decrease $\psi_{M\rm th}$ is to increase the laser beam energy. For a fixed $a_0^2 \tau/w$, a moderate $a_0$ (between 0.5 and 2.5 for most cases) can minimize $\psi_{M\rm th}$. Fig.~\ref{fig:a0_tau_scan} (c) and (d) show $\theta_{\rm opt}$ and $\psi_{M\rm th}$, respectively, as a function of $\tau/w$ for two values of $a_0$. It can be seen that both quantities tend to a well-defined value for $\tau/w > 5$, which suggests that further increasing the injection laser duration has a negligible effect. 

We have also performed a series of PIC simulations to determine the optimal angle $\theta_{\rm opt}$. 
In the simulations, the background plasma density is $n_p=4.9\times10^{16}\ \rm cm^{-3}$, thus $k_p^{-1}=24\ \rm \mu m$. The simulation box has a size of $10\times8\times8~k_p^{-3}$. The number of cells is $512\times2048\times256$ with 4 particle per cell for the plasma, and the time step is $3.8\times10^{-3}\omega_p^{-1}$ where $\omega_p=k_p c$ is the plasma frequency. The driver has a peak current of $1.84\ \rm kA$ (thus $\Lambda\approx0.2$), a peak density of $6\times 10^{17}\ \rm cm^{-3}$, a beam radius of $0.13~k_p^{-1}=3.12\ \rm \mu m$ (rms), a duration of $\omega_p^{-1} = 80\ \rm fs$ (rms) and an energy of 1 GeV. The laser frequency is $\omega_L=50\omega_p$. The optimal angle $\theta_{\rm opt}$ is plotted as circles with bars reflecting the uncertainty due to the finite $\theta$ step of the scan in Fig.~\ref{fig:a0_tau_scan} (c) for different values of $a_0$ and $\tau/w$, showing good agreement with the model prediction. 
For each configuration, the injection laser angle $\theta$ and its timing with respect to the driver have been scanned in order to determine $\theta_{\rm opt}$ to provide the highest amount of injected charge.

%\section{\label{sec:sim}The Injection Quantity and Quality}
\begin{figure}
	\begin{overpic}[width=0.48\textwidth]{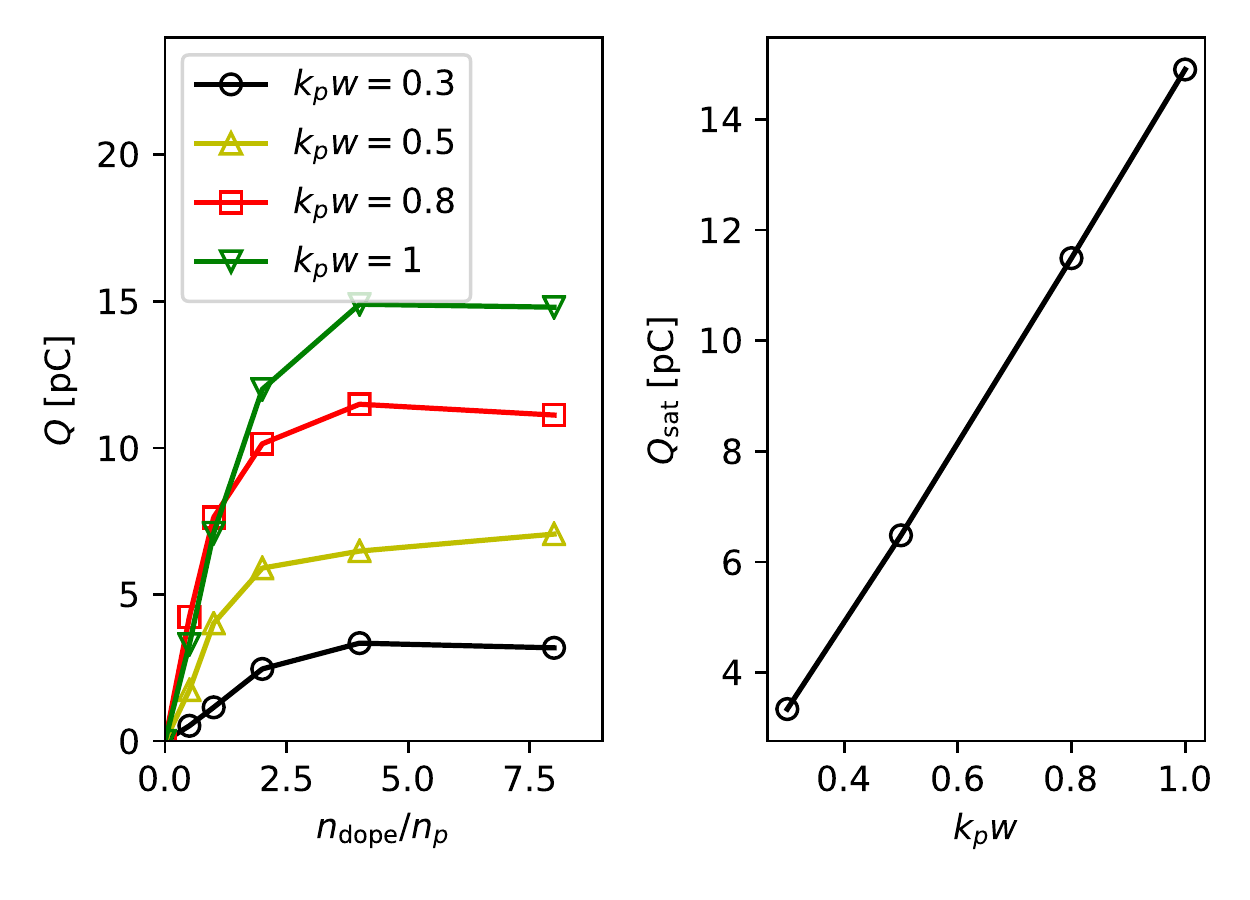}
        	\put(42,15){(a)}
        	\put(90,15){(b)}
	\end{overpic}
\caption{\label{fig:q_dope_w} PIC simulation results of trapped charge vs.\ doping ratio (a) and saturation charge vs.\ laser spot size (b) with an injection laser of $a_0=1$ and $\tau/w=2$. The calculation of charge is based on a plasma density of $n_p = 4.9\times 10^{16}\ \rm cm^{-3}$.}
\end{figure}
In order to analyze the amount of injected charge and the quality of the witness bunches generated by the proposed mechanism, we have performed PIC simulations with a similar configuration as in Fig.~\ref{fig:a0_tau_scan} (c). 
The laser features $a_0=1$, $\tau/w=2$ and $\theta = 73.8^\circ$ which is the optimal angle based on the models. The doping ratio $n_{\rm dope}/n_p$ and the laser waist radius are varied in the simulations. The timing of the injection laser is also adjusted to maximize charge for each case. The results are shown in Fig.~\ref{fig:q_dope_w}. In Fig.~\ref{fig:q_dope_w} (a), we plot the charge of the trapped bunch $70\ k_p^{-1}$ behind the injection z-position vs.\ the doping ratio $n_{\rm dope}/n_p$ for four different values of the laser waist radius $k_p w$. One can see that for small values of $n_{\rm dope}/n_p \lesssim 1$ the charge increases linearly with $n_{\rm dope}/n_p$, while for larger ratios the charge reaches saturation. The case $k_p w = 0.5$ and $n_{\rm dope}/n_p = 1$ was utilized as an example simulation in Fig.~\ref{fig:illu}. The saturation charge $Q_{\rm sat}$ vs.\ $k_p w$ is then plotted in Fig.~\ref{fig:q_dope_w} (b), which shows a linear dependency.

\begin{figure}
	\begin{overpic}[width=0.48\textwidth]{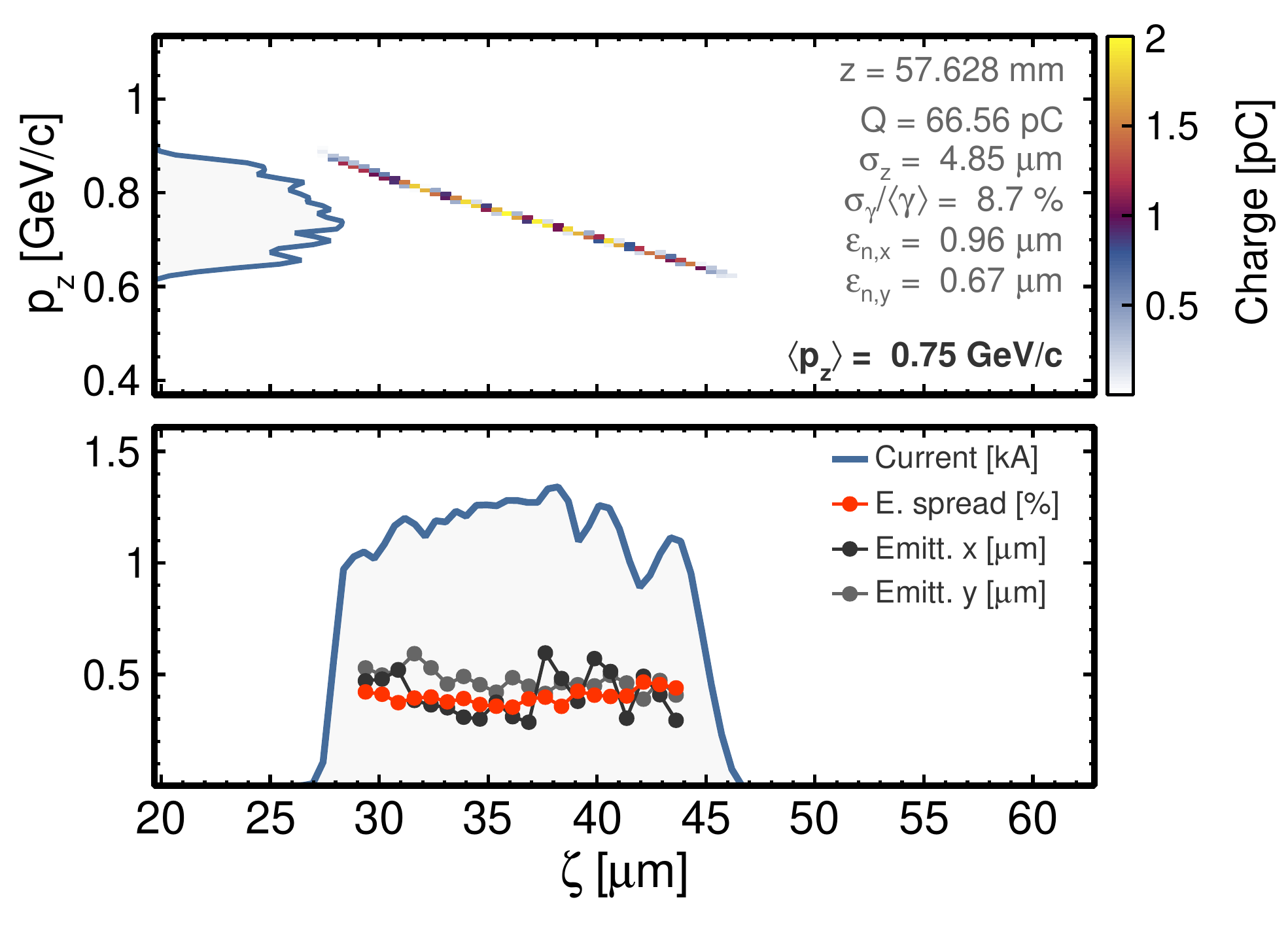}
	\end{overpic}
\caption{\label{fig:66pC} The longitudinal phase-space (top panel) and slice beam quality (bottom panel) of the witness beam for a case with increased driver current $I_{\rm peak} = 3\ \rm kA$, and other parameters being the same as the $k_p w=1$, $n_{\rm dope}/n_p=4$ case in Fig.~\ref{fig:q_dope_w}.}
\end{figure}

To further investigate the potential of this scheme, we ran simulations with slightly increased driver peak current $I_{\rm peak} = 3\ \rm kA$, so $\psi_M$ also increases but remains well below 1. All other simulation parameters remain the same with $w=k_p^{-1}=24\ \rm \mu m$, $n_{\rm dope}=4n_p$. The output beam data of an OSIRIS PIC simulation was applied as input to the quasi-static PIC code HiPACE~\cite{TMehrlingPPCF2014}, which allows for long distance simulations with largely reduced computational cost. The witness beam properties are shown in Fig.~\ref{fig:66pC}. At the acceleration distance of 57.6 mm (which is still far from driver-energy depletion), the witness beam gained $750~\rm MeV$ of energy, with an averaged sliced energy spread of $\sim0.4\%$ rms. The normalized emittance in the two transverse directions are $0.040~k_p^{-1} = 0.96~\rm mm \cdot mrad$ and $0.028~k_p^{-1} = 0.67~\rm mm \cdot mrad$. The current profile of the witness beam is approximately flat top with a maximum current of $1.3~\rm kA$ and $66~\rm pC$ in total. 

%\section{\label{sec:conclusion}Conclusion}
To conclude, a new injection scheme for PWFA was introduced which enables the trapping of high-quality witness beams even at moderate beam-driver currents. In this method, an assistive laser triggers the injection by the ponderomotive kick given to electrons released via ionization. The trapping condition in this scheme is discussed theoretically, and the optimal parameters for the injection laser are studied using numerical methods. PIC simulations are performed, demonstrating the generation of witness beams with sub-micrometer emittances and sub-percent uncorrelated energy spreads. 
In contrast to many ionization-based witness beam injection methods in a relativistic wakefield, this injection scheme works in the sub-relativistic regime. It provides a dark-current free environment and limits the injection to an extremely narrow temporal window. The initial phase-space volume of the trapped witness beam is largely constrained 
to a small fraction of the electrons scattered by the laser with the right momentum at the right position
resulting in low emittance witness beams.
Furthermore, the obtained witness beams feature a linear and negative energy-time correlation (chirp), which could be corrected by employing novel dechirping devices based on passive~\cite{RDArcyPRL2019, VShpakovPRL2019, YPWuPRL2019} or active~\cite{AFerranPousaPRL2019} plasma modules, to yield a final energy spread at the few per mille level. The hereby demonstrated witness beam quality is compliant with applications demanding high levels of beam brightness and low energy spread, such as free-electron lasers~\cite{FGrunerAPB2007}.  

\begin{acknowledgments}
We thank the OSIRIS consortium (IST/UCLA) for access to the OSIRIS code and acknowledge the use of the High-Performance Cluster (Maxwell) at DESY. We also gratefully acknowledge the Gauss Centre for Supercomputing e.V. (www.gauss-centre.eu) for funding this project by providing computing time through the John von Neumann Institute for Computing (NIC) on the GCS Supercomputer JUWELS at J{\"u}lich Supercomputing Centre (JSC). This work is supported by the Helmholtz MT ARD scheme and the Helmholtz ZT-0009 project.
\end{acknowledgments}

\bibliography{pdinj}% Produces the bibliography via BibTeX.

\end{document}